\documentclass[twocolumn,appendixfloats]{aastex62}
\usepackage{amssymb,amsmath}
\usepackage[cal=boondox]{mathalfa}
 
\newcommand{\footref}[1]{\textsuperscript{\ref{#1}}}
\newcommand{\simgt}{\lower.5ex\hbox{$\; \buildrel > \over \sim \;$}}
\newcommand{\simlt}{\lower.5ex\hbox{$\; \buildrel < \over \sim \;$}}

\def\m200{\mbox{$M_\mathrm{\rm 200m}$}}
\def\DSigma{\mbox{$\Delta\Sigma$}}
\def\Sigmacr{\mbox{$\Sigma_{\rm cr}$}}
\def\mhunit{\mbox{$\times10^{14}\,h^{-1}\,M_\sun$}}
\def\munit{\mbox{$10^{14}\,M_\sun$}}

\def\mpch{\mbox{$h^{-1}\,\mathrm{Mpc}$}}

\def\lcdm{\mbox{$\Lambda$CDM}}
\def\RM{\mbox{redMaPPer}}
\def\ha{\mbox{H$\alpha$}}

\def\o2{\mbox{OII}}
\def\ccf{\mbox{$\xi_{\rm g,cl}$}}
\def\acf{\mbox{$\xi_{\rm cl}$}}
\def\b{\mbox{$\mathcal{b}$}}
\def\bcross{\mbox{$\b^{\rm (cross)}$}}
\def\bauto{\mbox{$\b^{\rm (auto)}$}}
\def\bmlens{\mbox{$\b^{\rm 0}$}}
\def\blens{\mbox{$\b^{\rm 2h}$}}
\def\wp{\mbox{$w_{\rm p}$}}

\defcitealias{Medezinski2017}{M17}

\submitjournal{ApJ}
\shorttitle{\ha{} Cool-Core Assembly Bias}
\shortauthors{Medezinski et al.}

\begin{document}

\title{On the Assembly Bias of Cool Core Clusters  Traced by \ha{} Nebulae}

\author{Elinor Medezinski}
\affil{Department of Astrophysical Sciences, 4 Ivy Lane, Princeton University, Princeton, NJ 08544, USA}
\email{elinorm@astro.princeton.edu}

\author{Michael McDonald}
\affil{Kavli Institute for Astrophysics and Space Research, Massachusetts Institute of Technology, 77 Massachusetts Avenue, Cambridge, MA 02139, USA}

\author{Surhud More}
\affil{The Inter-University Center for Astronomy and Astrophysics, Post Bag 4, Ganeshkhind, Pune 411007, India}
\affil{Kavli Institute for the Physics and Mathematics of the Universe (WPI), Todai Institutes of Advanced Study, University of Tokyo, 5-1-5 Kashiwanoha,
Kashiwa 277-8583, Japan}

\author{Hironao Miyatake}
\affil{Institute for Advanced Research, Nagoya University, Nagoya 464-8601, Aichi, Japan}
\affil{Division of Particle and Astrophysical Science, Graduate School of Science, Nagoya University, Nagoya 464-8602, Aichi, Japan}
\affil{Kavli Institute for the Physics and Mathematics of the Universe (WPI), Todai Institutes of Advanced Study, University of Tokyo, 5-1-5 Kashiwanoha,
Kashiwa 277-8583, Japan}

\author{Nicholas Battaglia}
\affil{Department of Astronomy, Cornell University, Ithaca, NY, 14853, USA}


\author{Massimo Gaspari}
\affil{Department of Astrophysical Sciences, 4 Ivy Lane, Princeton University, Princeton, NJ 08544, USA}

\author{David Spergel} 
\affil{Center for Computational Astrophysics, Flatiron Institute, 162 5th Avenue, 10010, New York, NY, USA}
\affil{Department of Astrophysical Sciences, 4 Ivy Lane, Princeton University, Princeton, NJ 08544, USA}

\author{Renyue Cen}
\affil{Department of Astrophysical Sciences, 4 Ivy Lane, Princeton University, Princeton, NJ 08544, USA}

\begin{abstract}
Do cool-core (CC) and noncool-core (NCC) clusters live in different environments? We make novel use of H$\alpha$ emission lines in the central galaxies of redMaPPer clusters as proxies to construct large (1,000's) samples of CC and NCC clusters, and measure their relative assembly bias using both clustering and weak lensing. We increase the statistical significance of the bias measurements from clustering by cross-correlating the clusters with an external galaxy redshift catalog from the Sloan Digital Sky Survey III, the LOWZ sample.  Our cross-correlations can constrain assembly bias up to a statistical
uncertainty of 6\%. Given our H$\alpha$ criteria for CC and NCC, we find no significant differences in their clustering amplitude. Interpreting this difference as the absence of halo assembly bias, our results  rule out the possibility of  having different large-scale (tens of Mpc) environments as the source of diversity observed in cluster cores. 
Combined with recent observations of the overall mild evolution of CC and NCC properties, such as central density and CC fraction, this would suggest that either the cooling properties of the cluster core are determined early on solely by the  local ($<200$ kpc) gas properties at formation or that local merging leads to stochastic CC relaxation and disruption in a periodic way, preserving the average population properties over time. Studying the small-scale clustering in clusters at high redshift would help shed light on the exact scenario. 
\end{abstract}
 

\section{Introduction} 
\label{sec:intro}

In the modern picture of halo formation, clusters of galaxies, which are the last to collapse out of the large-scale structure (LSS) \citep{PressSchechter74}, grow in an inside-out manner in two growth phases \citep{Gunn1972}. In the early ``fast-rate" phase, rapid matter accumulation and major merger events build up the internal core of the cluster  inside a few times a characteristic scale radius ($r_s\approx 200$~kpc), erasing previous internal structure. In the subsequent ``slow-rate'' phase, the core is preserved, and the outskirts ($r > r_s$) gradually grow through moderate matter accretion. 
Thus, the internal structure of halos   contain signatures of their growth history \citep{Wechsler02,Zhao2003a,Ludlow2013,Correa2015}. Do the different baryonic properties of galaxy
clusters know or care about the different assembly histories, is a question worth investigating.

Interestingly, X-ray observations reveal that while on virial scales ($\sim$1~Mpc) clusters show remarkably self-similar entropy profiles as expected from  hierarchical formation, their cores  ($\lesssim200$~kpc) show a significant departure from self-similarity, with a variety of cooling phases \citep{Cavagnolo2009,McDonald2017,McDonald2018}.  
Cool-core (CC) clusters exhibit cuspy cores and low central temperatures and entropies \citep{Cavagnolo2008,Cavagnolo2009,Hudson2010}, whereas on the other end of the spectrum, non-cool-core (NCC) clusters  have disturbed cores with flatter central densities and high core entropies \citep[e.g.,][]{Ghirardini2019}. 
The brightest cluster galaxies (BCGs) in CCs  often coincide  with a `radio'-mode active galactic nuclei (AGN) (e.g., \citealt{Sun2009,McNamara2007,Hlavacek-Larrondo2015}), invoking a mechanical AGN feedback regulation \citep[see reviews by][]{McNamara2007,McNamara2012,Fabian2012,Gaspari2015,McDonald2018} to explain lower than expected ($\sim100$--$1000\,M_\odot\,$yr$^{-1}$) star formation rates observed in the core \citep[the ``cooling-flow problem"; see][]{Fabian1994,ODea2008}. Such AGN regulation cycle is tightly correlated with the ensemble warm/cold gas properties in CC clusters, such as high \ha/CO emission-line luminosity and significant velocity dispersions \citep[e.g.,][]{Donahue2000,Edge2001,Salome2003,McDonald2010,McDonald2012,Voit2015,Gaspari2018,Tremblay2018},  indicative of recent or ongoing star formation.

The formation mechanism leading to these differences in cluster cores is still
unclear. Following up  Sunyaev-Zel'dovich (SZ) \citep{Sunyaev1972}  detected
clusters in the X-ray with {\it Chandra}, \cite{McDonald2017} found little
evolution of the gas properties in the cluster cores since $z\sim1.6$,
suggesting that the core thermal equilibrium is established  early on, and remains
intact. Alternatively, transitions between CC and NCC may be periodic,
i.e., CC are formed and destroyed quickly or in equal numbers,  conserving a
constant population over time. If the cores of clusters are indeed preserved
over their lifetime,  it would imply that only the initial local central gas
density  dictates the in-situ  formation (or lack thereof) of a central AGN and
therefore the fate of the cluster core. In that case, the large-scale
environments of CCs should be indistinguishable from those of NCCs.

While high-resolution hydrodynamical simulations zooming in on the micro-
to meso-scale (0.1pc--100 kpc) physics have been successful in unveiling the
tight interplay between AGN feeding and feedback \citep[e.g.,][]{Gaspari2017,
Yang2019}, large-scale cosmological simulations ($>1$~Mpc) struggle to include
all the key physics in a self-consistent manner needed to retain predictive
power, leading to contrasting results depending on the chosen fine-tuned
parameters \citep[e.g.,][]{Rasia2015, Planelles2017, Barnes2018, Truong2018}.
As long as AGN feedback is implemented as a calibrated phenomenological subgrid model, robustly predicting the formation scenario of cluster cores in
cosmological simulations is likely to remain elusive.

In this case, can observations of galaxy clusters help us gain some insight
into understanding the physical processes that dictate the centers of galaxy
clusters? Dark matter halos are biased tracers of the underlying dark matter
distribution \citep[e.g.,][]{Kaiser1984, White1991}. This effect, especially on
galaxy cluster scales, depends  on halo mass to leading order. The higher the
mass, the larger the bias. However, at fixed mass, the assembly history of the
dark matter halo plays a significant role in setting
the halo bias \citep{Sheth2004, Gao2005, Wechsler06, Gao2007, Li2008,
Wang2011}, with late forming halos being more biased than their early forming
counterparts. Thus the differences in the halo bias as manifested by their
different clustering amplitudes could potentially be used as markers of
assembly history in order to learn about the physical processes going on in
galaxy clusters.

The assembly bias effect was first recognized in N-body simulations,
whereby e.g., massive ($>\munit$) halos that formed earlier showed lower clustering at fixed halo
mass \citep{Gao2005, Wechsler06, Gao2007, Wang2007}. Parameters other than halo
age were also found to correlate with clustering, such as concentration
\citep{Wechsler06, Gao2007, Faltenbacher2010, Villarreal2017}, spin
\citep{Gao2007, Lacerna2012, Lazeyras2017}, halo shape \citep{Lazeyras2017,
Villarreal2017}, and the level of halo substructure \citep{Wechsler06,
Gao2007}.  Most of these secondary properties correlate strongly with assembly
history, thus the effect was named assembly bias.  However, the secondary
parameters often also depend on the halo mass itself, making it hard to
disentangle the two effects \citep{Yang2006, Croton2007, Zentner2014,
Salcedo2018}. Partly for this reason, assembly bias has  been hard to confirm
observationally \citep{Lin2016}.  It is only marginally detected on galaxy
scales \citep{Montero-Dorta2017, Niemiec2018}, while on cluster scales,  where
the effect is predicted to be even weaker, observed results appear to suffer from
selection biases   \citep{Miyatake2016, More2016, Zu2017, Busch2017}.

In \citet[][hereafter M17]{Medezinski2017}, we developed a novel  methodology
to study the assembly bias of galaxy clusters. First, in order to increase the
statistical power, instead of using the two-point auto-correlation function
(ACF) of the clusters as typically done, we instead cross-correlate each target
cluster sample with a galaxy sample that probes the LSS with lower shot noise
\citep[see also][]{More2016}.  Subsequently, by comparing the cluster-galaxy
cross-correlation function (CCF) around CC and NCC clusters, one can study the
differences in the assembly histories of such galaxy clusters, and potentially
probe whether these two types of clusters are primordially distinct. In M17, we
used the X-ray core entropy, a rather expensive observable that can only be
resolved by the  {\it Chandra} X-ray satellite, in order to separate the CC/NCC
clusters. This restricted our sample size severely to only a few dozens of
clusters in each category. The measured bias therefore suffered from a
statistical error larger than the expected level of assembly bias.

With the prevalence of wide-field optical imaging surveys such as the Sloan Digital Sky
Survey \citep[SDSS;][]{Eisenstein2011},  samples of thousands of clusters can
now be used to study  cluster evolution statistically
\citep{Szabo2011,Soares-Santos2011,Wen2012,Rykoff2014,Oguri2014}. To date, the
most extensive public catalog applied to  SDSS/Data Release 8 (DR8) imaging,
the \RM{} catalog \citep{Rykoff2014,Rykoff2016} contains about 26,000 clusters.
Furthermore, about a million of the brightest galaxies were spectroscopically
followed up in  SDSS/DR12, including most of \RM{}'s BCGs.  As demonstrated by
\cite{McDonald2011b}, the signatures of cooling in a BCG spectrum  can be used
to distinguish between CC and NCC clusters and study their properties over
significantly larger ensembles than ever before.

In this paper, we aim to leverage the statistical power of the \RM{}
cluster sample, exploit the cooling information provided by the BCG
spectra, and apply the methodology developed in
\citetalias{Medezinski2017}, to statistically determine whether  CC
and NCC clusters have different clustering properties and thus come
from peaks collapsing from different initial conditions (i.e.,
assembly bias) or due to random processes (local in space and time) as
indicated by the inside-out formation model (i.e., no assembly bias).

This paper is organized as follows. In \autoref{sec:data} we present the
observational dataset used. In \autoref{sec:method} we review how to derive the
relative cluster bias from lensing and clustering. In \autoref{sec:results} we
present our results, utilizing  weak lensing  to disentangle
the mass bias effect, and presenting the relative bias as measured from
different lensing and clustering estimators. We discuss our results and compare with  recent simulations in \autoref{sec:discussion} and summarize and conclude in \autoref{sec:summary}. Throughout the paper, we  adopt a \lcdm{} cosmological model, where $\Omega_m=0.27$, $\Omega_\Lambda=0.73$, and $h = H_0/100\,{\rm km\,s^{-1}\, Mpc}^{-1} = 1$. Unless otherwise stated, we quote median and 68\% confidence interval values.

\section{Data}
\label{sec:data}
Similar to the methodology presented in \citetalias{Medezinski2017}, our goal
is to compare two subsamples of clusters that differ in their cooling state,
having either cool or non-cool cores, and test if they have different
clustering amplitudes. To do so, we measure the clustering of galaxies as
tracers of the large-scale structure around each cluster subsample. The ratio
of their clustering gives an estimate of their {\it relative} halo bias
(see \autoref{sec:Xmethod} for definitions).  In this paper, we increase the
statistical power of the method in \citetalias{Medezinski2017} by using a
larger sample drawn from the \RM{} catalog, and make novel use of the intensity
of cooling lines in  BCGs to differentiate between CC and NCC
clusters.  In the following section we describe the construction of the
cluster subsamples and present the correlation between BCG emission line
luminosity and cluster core entropy. We then briefly describe the galaxy
catalog used for the cross-correlation study, LOWZ.

\subsection{The \RM{} Cluster Catalog}
\label{subsec:RM}

\begin{figure}[t]
\includegraphics[width=0.5\textwidth,clip]{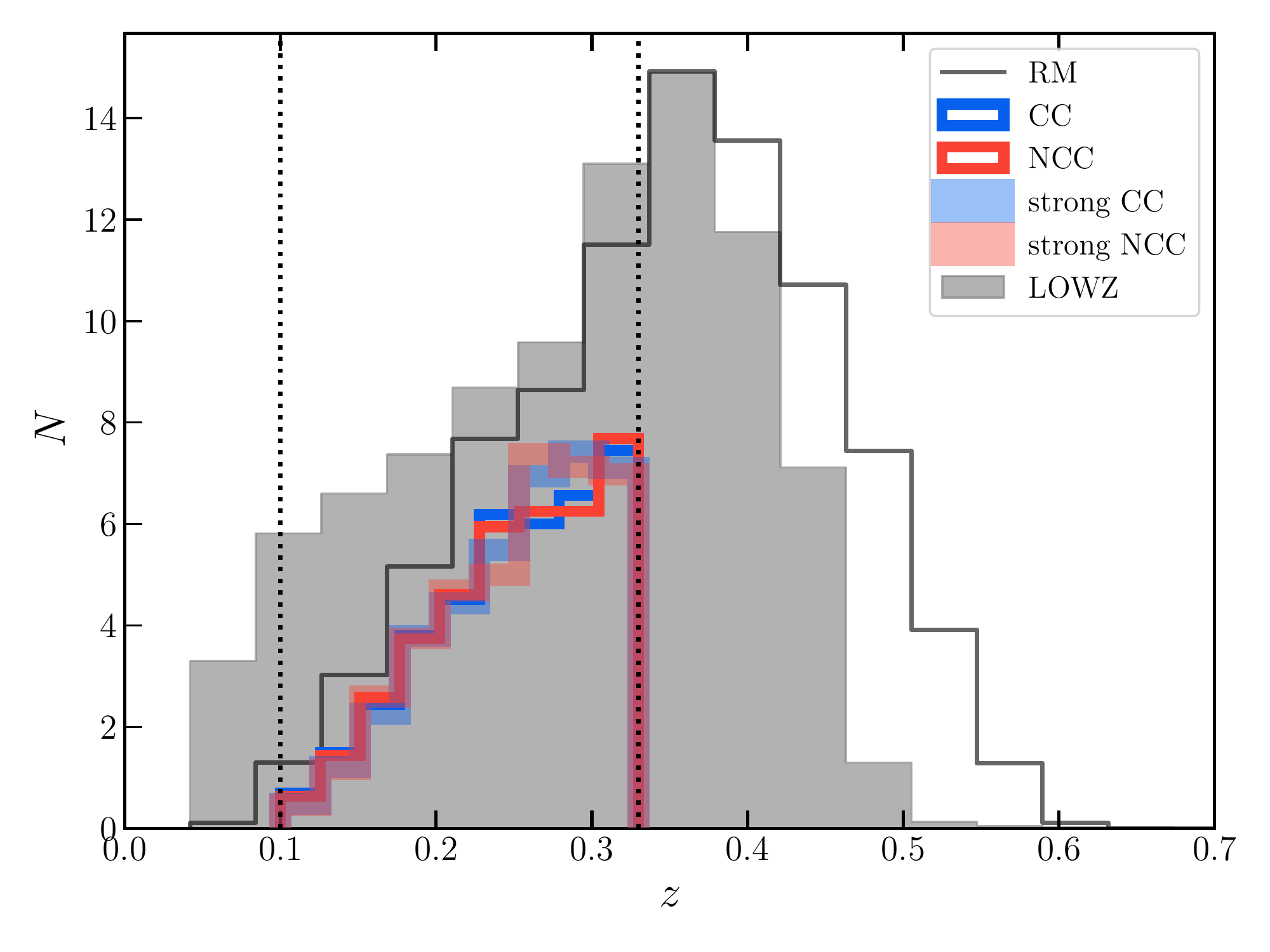}
\caption{Redshift distribution of the different samples: the full parent cluster sample, \RM{} (black line), CC and NCC samples (thick blue and red lines, respectively),  strong CC and NCC (thin blue and red lines, respectively), and the LOWZ galaxy sample (gray). All samples have arbitrary normalization. }
\label{fig:zhist}
\end{figure}

\begin{figure}[h]
\includegraphics[width=0.5\textwidth,clip]{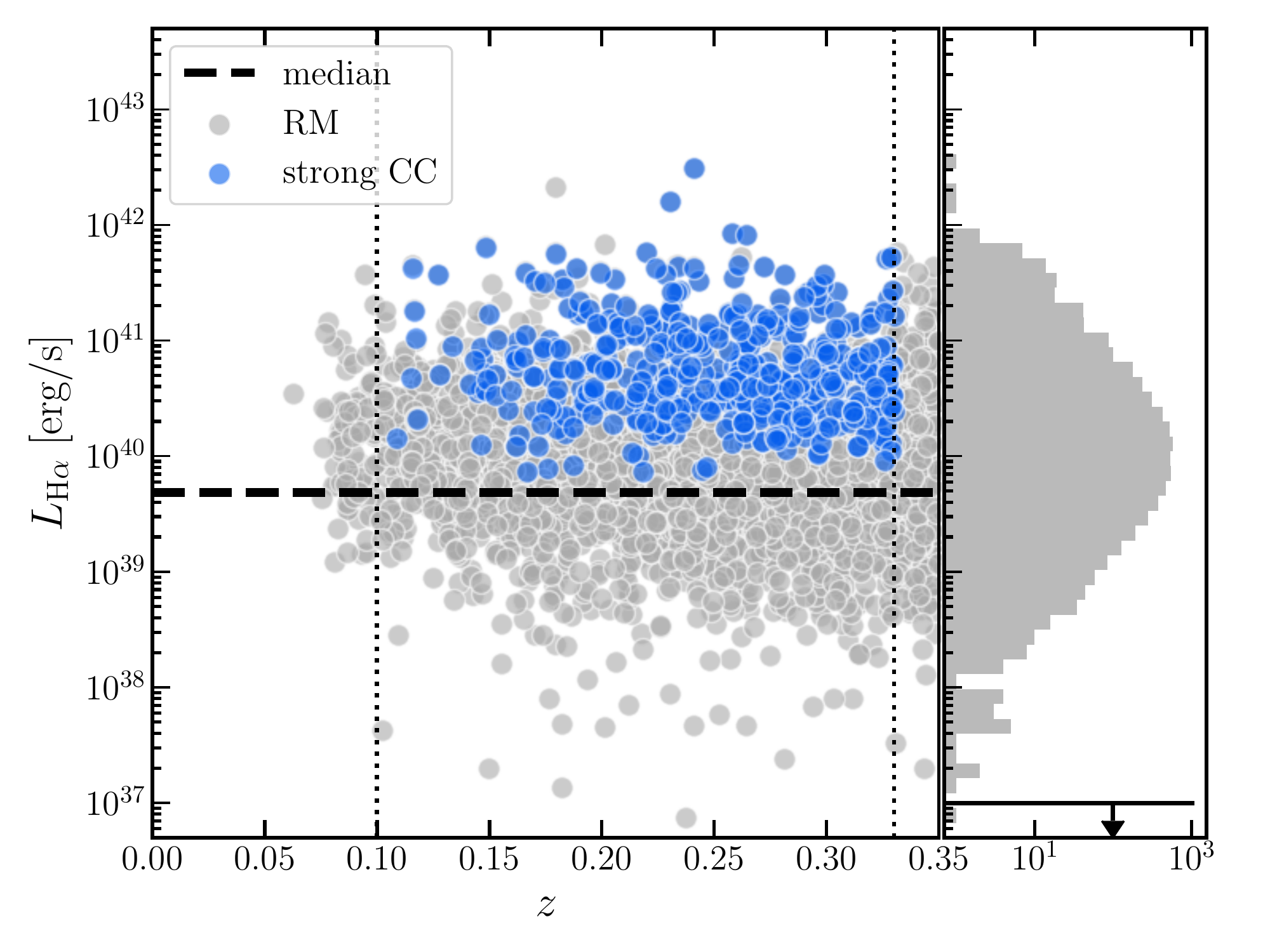}
\caption{\ha{} luminosity for each \RM{} BCG with robustly measured spectroscopy as a function of redshift. The redshift limits adopted are marked as vertical dotted lines.
Dashed black line indicates the median luminosity threshold above (below)  which clusters are considered CC (NCC) clusters.
We also mark clusters which are considered strong CC (blue points), for which the \ha{} luminosity detection is significant (i.e., AoN$>2$; see \autoref{subsubsec:samples}). Equivalently, those with no \ha{} emission line detected are considered strong NCC (not marked on this logarithmic scatter plot, but  indicated by the black upper limit drawn on the histogram to the right). 
}
\label{fig:Lz}
\end{figure}
\begin{figure}[h]
\includegraphics[width=0.5\textwidth,clip]{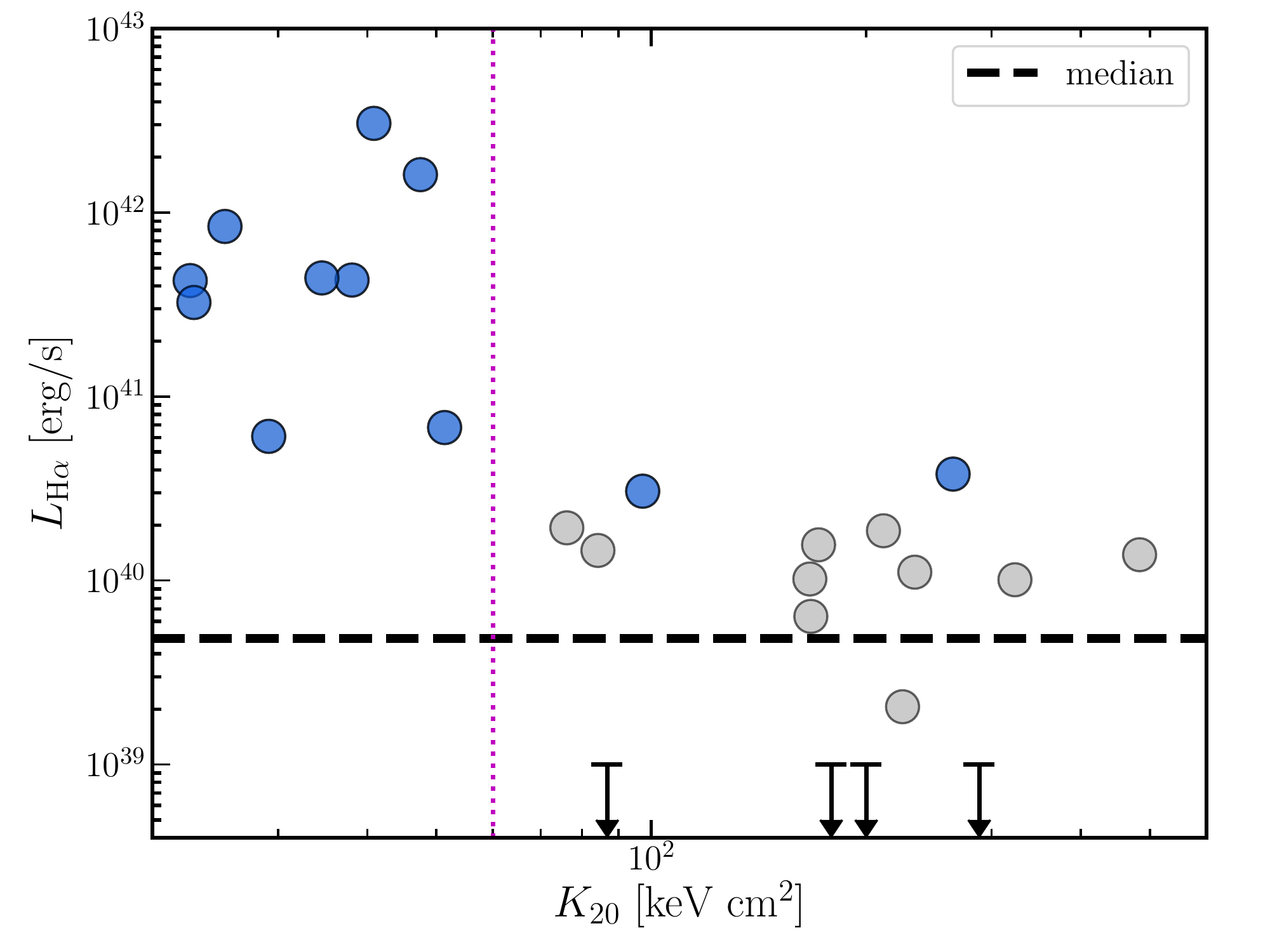}
\caption{\ha{} luminosity  as a function of central entropy inside 20~kpc, $K_{20}$, for each \RM{} BCG  having both spectra from BOSS and entropy measured from the ACCEPT ({\it Chandra}) database. 
Upper limits are drawn for BCGs with no \ha{} emission lines. We mark  the median luminosity level (dashed black line) we use to separate CC from NCC,  as in \autoref{fig:Lz}. The entropy threshold of $K_{20}=60$~keV~cm$^2$ used in \citetalias{Medezinski2017} to separate CC and NCC clusters is marked as vertical dotted magenta line. Strong CC clusters are indicated as blue circles. Evidently, all clusters previously considered as CC by the entropy threshold are indeed strong CC clusters using the \ha{} definition.}
\label{fig:L-K20}
\end{figure}

We use the latest optically selected galaxy cluster catalog detected from
SDSS/DR8 with the \RM{} cluster finder algorithm version 6.3 \citep{Rykoff2014,
Rykoff2016} available online\footnote{\label{note1}\url{http://risa.stanford.edu/redmapper/}}.  For each cluster, the catalog lists redshift, richness estimate $\lambda$ (approximately the number
of cluster galaxies above $0.2\,L_*$) and a position, where the position is
that of the BCG. BCG identification in \RM{} is good to 85\% \citep{Rykoff2014,Rozo2014a,Hoshino2015}. This version of the catalog
contains a total of 26,111 clusters detected in $\sim10,000$~deg$^2$, spanning
redshifts in the range $0.08<z<0.6$ and richnesses $\lambda>20$. We present the redshift distribution of the full cluster sample in \autoref{fig:zhist}
(black), where throughout we use the spectroscopic redshift from SDSS (see item 3 in \autoref{subsubsec:ELines}). A cluster random catalog\footref{note1} has also been constructed \citep{Rykoff2016},
appropriate for  large-scale
two-point correlation studies such as the one conducted here. We use weights  provides in the random catalog to account for survey depth and redshift completeness.
Since we aim to cross-correlate the clusters  with the LOWZ galaxy catalog, both catalogs need
to span the same spatial region on the sky. For LOWZ, some patches observed
early on in the survey were masked due to a bug in the initial targeting
software \citep[see][]{Reid2016}. We therefore apply the LOWZ North and South
masks onto the \RM{} cluster and random catalog. Finally, we confine our
analysis to clusters within $0.1<z<0.33$ for an approximately volume-limited
sample \citep{Miyatake2016}, and apply the same redshift limits to the random
sample.

\subsubsection{Emission-line luminosity as a cool-core indicator}
\label{subsubsec:ELines}
In order to determine if a cluster is in CC or NCC phase, we search for
signatures of cooling in the spectrum of its BCG. It
has been shown that  clusters that  harbor multiphase gas with a cool gas core
and low central entropy ($\lesssim30$~keV~cm$^2$) typically also exhibit strong
emission-line luminosities from \ha{} filaments of gas cooling onto their BCG
\citep[e.g.,][]{Cavagnolo2008,McDonald2011b,Gaspari2018}. We therefore calculate \ha{}
emission-line luminosities for each \RM{} BCG following these steps:
\begin{enumerate}
\item We query emission-line fluxes from the emissionLinesPort\footnote{\url{http://skyserver.sdss.org/dr12/en/help/browser/browser.aspx\#\&\&history=description+emissionLinesPort+U}} 
table \citep{Thomas2013} in SDSS/DR12 for all galaxies within the same redshift range as the clusters.
\item Where there are duplicate spectra per galaxy  (i.e., measured by both SDSS and BOSS), we use the most recent one (BOSS).
\item We cross-match the \RM{} catalog with the emission line table using the SDSS {\sc objid} identifier. This provides emission line flux estimates (and spectroscopic redshift) for each cluster. About 17,000 (out of the 26,000) are left after this matching.
\item We remove entries where no \ha{} continuum was measured (i.e., Flux\_Cont\_Ha\_6562=$-9999$ or Flux\_Cont\_Ha\_6562\_Err=0)
 or measured with low signal-to-noise, S/N$<5$\footnote{S/N is determined as the continuum flux over its error measured at \ha{}  S/N = Flux\_Cont\_Ha\_6562/Flux\_Cont\_Ha\_6562\_Err.}\footnote{Only 6 clusters are removed by this cut. Most clusters have S/N$>60$.}.
\item We convert flux to luminosity using the base cosmology and the BCG spectroscopic redshift.
\item For low redshift galaxies ($z<0.3$), the  aperture size of the spectral fiber can be smaller than the galaxy size, causing the luminosity to be underestimated. Following \cite{McDonald2011b}, we correct for this effect by assuming no redshift evolution of the emission-line luminosity, and fit  a power-law model to $L(z)$. Since the fiber size changed from  $3\arcsec$ in the SDSS phase to $2\arcsec$ in the BOSS phase, we calculate the correction separately for SDSS and BOSS spectra (determined by plate number). 
\end{enumerate}
In total, we are left with 6,687 clusters with emission-line information on their BCG in the redshift range $0.1<z<0.33$ with  LOWZ spatial coverage. The main limiting factor is the redshift range we adopt.

We present the corrected \ha{} luminosity as a function of redshift in \autoref{fig:Lz} for all the BCGs in the \RM{} sample. 
Many ($\sim30\%$) BCGs have emission-line luminosities below the tail of the distribution, $L_{\ha}\lesssim10^{38}$~erg/s, with nearly all having no \ha{} detection (indicted as an upper limit on the histogram).

Next, we examine the correlation of emission-line luminosity with central entropy. We use the ACCEPT sample by \cite{Cavagnolo2009} who measured entropy profiles for 241 clusters observed with Chandra. We cross-match between \RM{} and the ACCEPT catalog within a $0.5\arcmin$ aperture, resulting in 30 clusters.  In \autoref{fig:L-K20} we plot the \ha{} emission-line luminosity as a function of central entropy, $K_{20}$, defined as the mean entropy inside 20~kpc. 
Clusters with no \ha{} detection (i.e., Flux\_Ha\_6562=0)  are indicated as
upper limits (arbitrary-level barred arrows; no errorbars are given for these null measurements). As can be seen from the figure, the luminosities
are anti-correlated with the central entropy, as previously reported.
Specifically, below $K_{20}=60$~keV~cm$^{-2}$ (pink dotted vertical line), the
threshold we have used in \citetalias{Medezinski2017} to separate CC from NCC,
all clusters have significantly higher luminosities, indicative of stronger
cooling flows onto their BCG.

\subsubsection{CC and NCC sample definitions}
\label{subsubsec:samples}
We now use the \ha{} luminosity information to create CC and NCC subsamples from the parent \RM{} sample. First, we divide the sample evenly by the median luminosity value
indicated by the dashed black line 
 in both \autoref{fig:Lz} and \autoref{fig:L-K20}. All clusters above this value are considered CC, while those below are considered NCC. However, selecting CCs based on  \ha{} lines introduced a bias, since such features can be better determined at lower redshifts. This makes the redshift distribution of CCs skewed toward lower redshift with respect to that of NCCs. For this reason, we match the redshift distributions of the CC and NCC samples by downsampling both samples. This reduced the sample sizes  further by $\sim10\%$ to about 3,000 clusters in each. The redshift distributions of CC (blue) and NCC (red) subsamples after this procedure are presented in \autoref{fig:zhist}. Their mean redshift is $\langle z \rangle = 0.25\pm0.06$.

As noted in the  emissionLinePort documentation\footnote{\url{https://www.sdss.org/dr12/spectro/galaxy_portsmouth/}}, some fits to the spectra do not yield significant emission-line flux measurements. Only those with amplitude-over-noise (AoN) larger than two are considered significant emission-line fluxes. We therefore select another more restrictive subset, ``strong CC", of clusters with significant \ha{} flux (AoN\_Ha\_6562$>2$) totaling 485 clusters\footnote{\label{note2}The strong CC and strong NCC redshift distributions are also matched, as for the CC/NCC samples}. The strong CC sample is presented in \autoref{fig:Lz} and in \autoref{fig:L-K20} (matched with ACCEPT) as blue circles. As can be seen from \autoref{fig:L-K20}, all 9 clusters below our previous entropy-based definition ($K_{20}<60$~keV~cm$^2$) are considered strong CC based on the new \ha{} definition, and only two strong CC clusters are considered NCC based on their entropy. Raising the AoN threshold further to exclude those would cut the strong CC sample in half ($\sim$280 clusters), too small for a statistically significant analysis. The

We similarly select a restrictive ``strong NCC" subset of clusters with no \ha{} line detection (i.e., Flux\_Ha\_6562=0), totaling 1,778 clusters\footref{note2}.
The redshift distributions of the strong CC and NCC\footref{note2} are also indicated in \autoref{fig:zhist} as thick transparent blue and red lines, respectively. Their mean redshifts are the same as for their parent CC/NCC samples. We summarize the basic properties of all the cluster subsamples in \autoref{tab:samples}.

One limitation of the \RM{} algorithm is that it a-priori selects ``red'' galaxies as cluster members. For this reason, strong CC clusters with extreme star-formation in their BCGs leading to bluer colors may be missing from this catalog. \cite{Rykoff2014} show examples of known CC clusters that are still detected, but their BCG is misidentified, leading to larger miscentering. In our above classification, such misidentification may lead to assignment of a CC cluster as NCC. However, \cite{Rykoff2014} demonstrate that such catastrophic miscentering happens for roughly $\lesssim5\%$ of all clusters. 
We therefore do not expect a large contamination of the NCC sample by CCs. However, as also indicated from the final sizes of our strong CC/NCC samples, this selection effect does diminish the number of strong CCs in the catalog.

\subsubsection{Richness as a mass proxy}
\label{subsec:richness}
\begin{figure}[t]
\includegraphics[width=0.5\textwidth,clip]{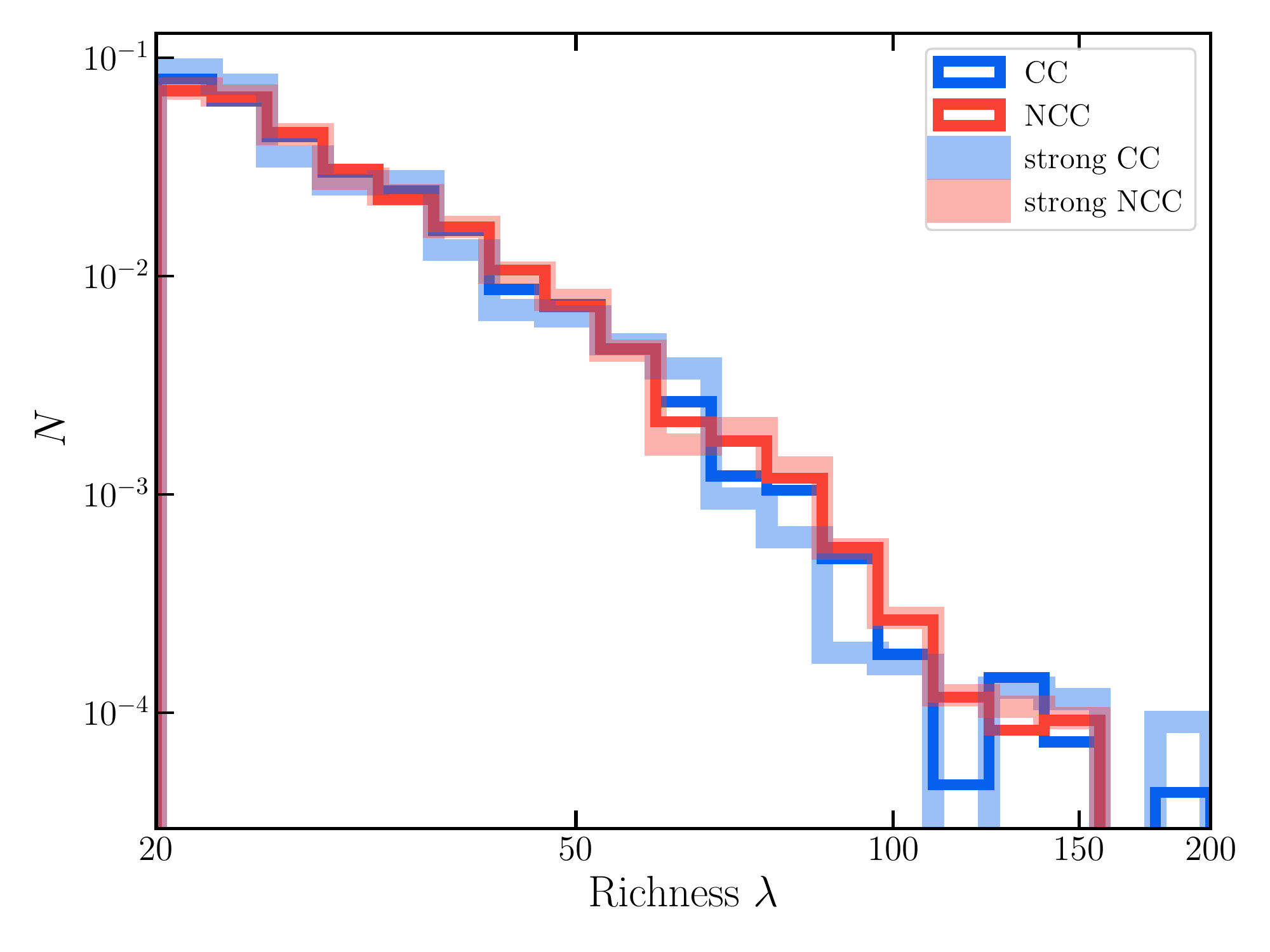}
\caption{Normalized richness distribution of the different CC and NCC samples, as indicated in the legend. 
}
\label{fig:Lhist}
\end{figure}

The bias of halos on galaxy cluster scales depends first and foremost on mass.
Thus, it is important to ensure that the two cluster subsamples we compare have
comparable mean mass before exploring any secondary dependencies on other
properties.
The \RM~ catalog contains a richness measurement for each cluster, which can be
considered as a mass proxy. Although richness is a very noisy mass proxy, our
samples are large enough that the error on the mean mass of each sample is
small. In the \autoref{app:null}, we will explore the level of uncertainty in
the use of richness as a mass proxy by randomly subsampling from the \RM{}
sample. We present the richness distribution for the CC, NCC, strong CC and
strong NCC samples  in \autoref{fig:Lhist}. The samples have consistent
distributions, and their mean richness (and therefore, mass) are consistent
within the errors. We list the mean mass based on the
mass-richness relation from \citep{Simet2017} in \autoref{tab:samples}. 

We will further explore the mean masses of each cluster subsample using
weak gravitational lensing in \autoref{subsec:WL}. 

\begin{deluxetable}{lrccc}
\tablecaption{Galaxy and  Cluster Samples Properties\label{tab:samples} }
\tablewidth{0pt}
\tablehead{
\colhead{Name} & \colhead{$N$} & \colhead{$\log\langle L_{\ha} \rangle$} & \colhead{${M^\lambda_{200\rm m}}$} & \colhead{$M^{\rm lens}_{\rm200m}$} \\
& \colhead{} &\colhead{[ergs/s]} & \colhead{\mhunit} &\colhead{\mhunit} 
}
\startdata
LOWZ & 239904  &  &&\\
\RM{} &  6687 &  & &\\ 
\hline
CC  & 3053 & 40.5 & 1.88$\pm$0.03 & $1.99^{+0.21}_{-0.19}$ \\
NCC& 3035 & 39.0 & 1.90$\pm$0.03 & $2.10^{+0.18}_{-0.16}$\\
strong CC  & 485 & 41.0 & 1.83$\pm$0.07 & $1.99^{+0.45}_{-0.40}$\\
strong NCC&1778 & --- & 1.91$\pm$0.04 & $2.11^{+0.23}_{-0.21}$\\
\enddata
\tablecomments{Strong NCC are defined as clusters whose BCG has no \ha{} emission line detection, therefore no luminosity is indicated.
}
\end{deluxetable}

\subsection{LOWZ galaxy sample}
\label{subsec:LOWZ}
We use the LOWZ spectroscopic-redshift galaxy catalog
\footnote{\url{https://data.sdss.org/sas/dr12/boss/lss/}} \citep{Reid2016},
which is drawn from the Baryon Oscillation Spectroscopic Survey
\citep[BOSS;][]{Dawson2013}. BOSS is part of SDSS DR12 \citep{Alam2015}. 
BOSS covers a total effective area of 8,337 deg$^2$, and the LOWZ sample contains 463,044 galaxies with spectroscopic redshifts \citep{Reid2016}. 
We limit both the galaxy catalog and the corresponding random catalog to the
redshift range set by the \RM{} catalog, $0.1<z<0.33$. 
The basic properties of the cluster and
galaxy samples we use in the cross-correlation analysis are listed in
Table~\ref{tab:samples}.


\section{Methods}
\label{sec:method}

In this section we detail the two independent methodologies we use for determining the level of assembly bias: clustering and weak lensing.

\subsection{Galaxy Clustering}
\label{sec:Xmethod}

Here we review the methodology developed in \citetalias{Medezinski2017}, cross-correlating a cluster sample with a larger galaxy sample in order to improve the statistical inference. The full details are given in  \citetalias{Medezinski2017}, and so we only briefly summarize here.

The simple linear, deterministic galaxy bias relates between the galaxy overdensity, $\delta_\mathrm{g}(x)$, and underlying matter overdensity, $\delta(x)$ \citep{Kaiser1984}, 
\begin{equation}\label{eq:delta}
\delta_\mathrm{g}(x) = b_\mathrm{g} \delta(x)
\end{equation}
where overdensity is defined with respect to the mean density, $\delta(x)
\equiv \rho(x)/\bar{\rho}-1$. In practice, we make use of the two-point
correlation function, $\xi(r) \equiv \langle\delta(x)\delta(x-r)\rangle$
\citep{Peebles1973,Peebles80}. Under the above assumptions,  the galaxy bias
relates the galaxy two-point auto-correlation to the underlying matter
correlation function, such that,
\begin{equation}
\xi_{\rm g}(r) = b_\mathrm{g}^2 \xi_{\rm m}(r).
\end{equation}
Spectroscopic observations of galaxies typically allow us to measure the galaxy
correlation function. In the linear, deterministic galaxy bias model, it is
proportional to the correlation function of the underlying matter distribution, described
by a constant factor. 

One can alternatively cross-correlate different samples of galaxies, or
galaxies and clusters as in our case. For the galaxy-cluster CCF, the above
model yields
\begin{equation}\label{eq:ccf}
\ccf(r) = b_{\rm g} b_{\rm cl} \xi_{\rm m}(r)\,,
\end{equation}
where $b_{\rm g}$ and $b_{\rm cl}$ denote the galaxy and the cluster bias,
respectively. Since we correlate each of the two cluster samples (CC, NCC)
with the same galaxy sample (LOWZ in our case), the ratio of these two
cross-correlations simply traces the {\it relative} bias of NCC clusters with
respect to CC clusters,
\begin{equation}\label{eq:bias_cross}
\bcross(r)\equiv b_{\rm NCC}/b_{\rm CC} = \frac{\xi_{\rm g, NCC}(r)}{\xi_{\rm g,CC}(r)}\,.
\end{equation}
The galaxy bias term, $b_\textrm{g}$, automatically cancels out. Note that
using clustering alone  (without assuming a halo model) one cannot  constrain
the individual cluster biases, $b_{\rm CC},~b_{\rm NCC}$, without assuming a
cosmological model. We can instead constrain their ratio, the {\it relative}
bias,  defined as $\b$, and the superscript notation (cross) indicates the use
of CCFs in the measurement.

Similarly, the ACF of clusters, $\xi_{\rm cl}(r)$, would simply be
\begin{equation}\label{eq:acf}
\acf(r) = b_{\rm cl}^2 \xi_{\rm m}(r),
\end{equation}
so that the relative bias can be derived from the CC and NCC auto-correlations as,
\begin{equation} \label{eq:bias_auto}
\bauto(r) =  \sqrt{ \frac{\xi_{\rm NCC}(r)}{\xi_{\rm CC}(r)}. }
\end{equation}
The disadvantage of the ACF is that it is measured with a lower
statistical precision compared to the CCF, since it requires the use of the
smaller cluster sample.

\subsection{Weak Lensing}
\label{sec:WLmethod}

Weak gravitational lensing (WL) induces a coherent tangential distortion to
the shapes of background galaxies, proportional to the underlying halo excess
surface mass density profile of the lensing cluster. The WL signal is
related to the cluster-matter cross-correlation function and therefore allows
us to determine both the halo total mass (to validate the CC/NCC samples have
similar masses), and also independently infer the linear bias parameter from
the larger scales of the lensing profile.

The WL methodology as applied to SDSS has been extensively reviewed in the
literature \citep[e.g.,][]{Mandelbaum2005,Mandelbaum2013,Simet2017,Murata2018},
so we only briefly summarize it here. We estimate the mean projected cluster
mass density excess profile $\DSigma(r)$ by stacking the shear (as measured
from the ellipticities) of source galaxies $s$ over multiple clusters $l$ that
lie within a given cluster-centric radial annulus $r$,
\begin{equation}\label{eq:DSigma}
\DSigma(r) = B(r) \frac{1}{2\mathcal{R}}\frac{ \sum\limits_{l,s} w_{ls} e_{t,ls}\Sigmacr_{,ls} } { \sum\limits_{l,s}w_{ls}},
\end{equation}
where the double summation is over all clusters and over all sources associated
with each cluster (i.e., lens-source pairs).  $\Sigmacr =  \frac{c^2}{4\pi
G}\frac{D_A(z_s)}{D_A(z_l)D_A(z_l,z_s)(1+z_l)^2},$  is the  critical surface
mass density, where $G$ is the gravitational constant, $c$ is the speed of
light, $z_l$ and $z_s$ are the lens and source redshifts, respectively, and
$D_A(z_l)$, $D_A(z_s)$,   and $D_A(z_l,z_s)$ are the angular diameter distances
to the lens, the source, and the lens-to-source, respectively. The extra factor
of  $(1+z_l)^2$ comes from our use of comoving coordinates
\citep{Bartelmann2001a}. The photometric redshifts of source galaxies were
estimated with  {\sc zebra} and a photometric redshift bias correction was
applied \citep{Nakajima2012}.  The minimum variance estimator requires the
weights to be $w_{ls} = \frac{1}{\Sigmacr_{,ls}^2}
\frac{1}{\sigma_{e,s}^2+e_{{\rm rms}}^2}$, where $\sigma_e$ is the shape
measurement uncertainty due to pixel noise, and $e_{\rm rms}=0.365$ is the
intrinsic shape noise. The  `shear responsivity' factor, ${2\mathcal{R}}$,
represents the response of the ellipticity, $e$, to a small shear
\citep{Kaiser1995,Bernstein2002}.
The factor $B(r)$, corresponds for the boost factor, needed to correct for the
dilution effect, which arises from contamination by unlensed cluster galaxies
having imperfect photometric redshift estimates. The boost is estimated by
comparing the weighted number density of source-lens pairs to that around
randoms points, $B(r) = \frac{N_r}{N_l}\frac{\sum_{ls} w_{ls}}{\sum_{rs}
w_{rs}}$.

Following \cite{Miyatake2016}, we fit each lensing profile with a five-parameter model,
\begin{equation}
\begin{split}
\Delta\Sigma&(r; \m200,c_{200\rm m},q_{\rm cen},\alpha_{\rm off},b) = \\
& q_{\rm cen}\Delta\Sigma^{\rm NFW}(r; \m200,c_{200\rm m})\\
&+(1-q_{\rm cen})\Delta\Sigma^{\rm NFW,off}(r; \m200,c_{200\rm m},\alpha_{\rm off})\\
&+\Delta\Sigma^{\rm 2-halo}(r; b).
\end{split}
\end{equation}
The first term arises from the halo mass profile for a fraction $q_{\rm cen}$ of
clusters whose BCGs as identified by \RM{} represent the true cluster
centers, while the second term describes the profile of the off-centered
clusters. We assume that the offsets follow a Gaussian distribution in three
dimensions, with  $u_{\rm off} \propto \exp{\left[-r^2/(2\alpha_{\rm
off}^2r_{200\rm m}^2)\right]}$, where $\alpha_{\rm off}$ describes the ratio of
the off-centering radius to $r_{200\rm m}$, the radius at which the enclosed
mass density is 200 times the mean density of the universe. For both components
we adopt the smoothly-truncated NFW \citep*{NFW96} model \citep{Oguri2011}.
The final term models the contribution from the surrounding large-scale
structure, i.e., the two-halo term. We employ the model given as
$\Delta\Sigma^{\rm 2-halo}(r; b) = b\int{k{\rm d} k/(2\pi) \bar{\rho}_m P^{\rm
L}_{\rm m}(k;z_{\rm cl})J_2(kr)}$, where  $\bar{\rho}_m$ is the mean mass
density today, $b$ is the linear bias parameter, and $P^{\rm L}_{\rm
m}(k;z_{\rm cl})$ is the linear mass power spectrum at the averaged cluster
redshift $z_{\rm cl}=0.25$, for the
\lcdm{} model.
We can then directly probe the relative bias from the individual bias parameters fitted to the lensing profile, defined as
\begin{equation}\label{eq:blens}
\blens \equiv \frac{b_{\rm NCC}}{b_{\rm CC}}.
\end{equation}

As a baseline expectation, we can use the halo bias model of
\citep{Tinker2010}, calibrated from numerical simulations, in order to estimate
the {\it expected} halo bias level for each sample based on the derived mean WL
mass. Under the zeroth order assumption of no assembly bias, we can then define
the expected relative bias as,
\begin{equation}\label{eq:bmlens}
\bmlens \equiv \frac{b(M_{\rm NCC})}{b(M_{\rm CC})}.
\end{equation}
Any deviation from this value should therefore give an estimate of the level of assembly bias. 
Specifically, we define $f$ as the ratio of the measured bias from each methodology defined above (e.g., \bcross, \bauto, \blens) to the expected mass-only bias, \bmlens, such that our model for the measured bias is 
\begin{equation}\label{eq:f}
 \b = f \bmlens.
\end{equation}
Confirmation of assembly bias requires $f\neq1$. 
We will therefore obtain the marginalized
posterior distribution of $f$ from fitting the above model to each measured relative bias. We perform the fitting with the public code {\sc emcee} \citep{Foreman-Mackey2013}.

\section{Results}
\label{sec:results}

In this section we present the  WL and clustering analyses of the CC and NCC cluster samples defined above and the  mean relative bias we derive for each methodology.

\subsection{Weak Lensing}
\label{subsec:WL}

\begin{figure*}[t]
\includegraphics[width=\textwidth,clip]{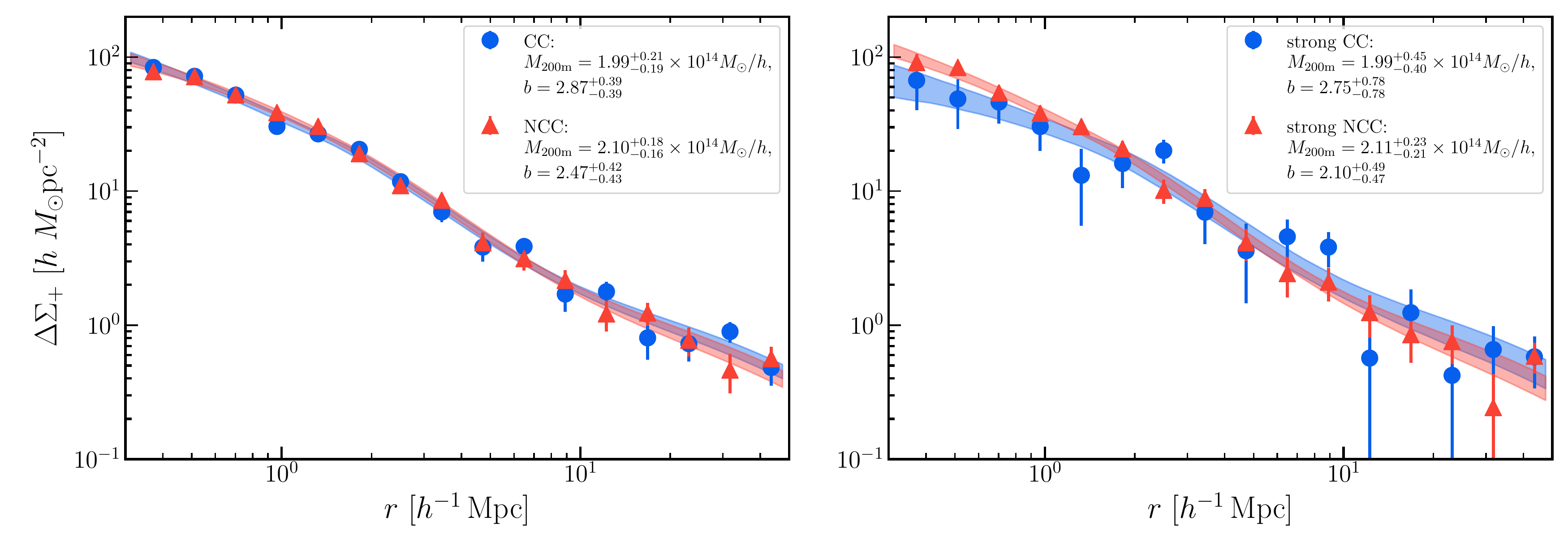}
\caption{Stacked cluster WL excess surface mass density profiles for the CC and NCC samples (left), and the strong CC/NCC samples (right). Mean masses and halo bias values derived from model fitting the WL profiles are given next to each set. The NCC mean mass is about $1\sigma$ above the CC mass (see \autoref{tab:samples}), whereas the CC galaxy bias is about $1\sigma$ above the NCC galaxy bias (and similarly for the strong CC/strong NCC samples). }
\label{fig:Sigma}
\end{figure*}

Here we make use of lensing data to first determine the mean masses of the CC and NCC cluster samples and test if they are indeed comparable as indicated by the mass-richness scaling relation.  Subsequently, cluster WL profiles will be used to determine the level of expected mass bias and the level of measured bias.

We use the SDSS/DR8 shape catalog \citep{Reyes2012,Mandelbaum2013} measured with the reGaussianization technique \citep{Hirata2003}. The  systematic uncertainties on shape measurements have been thoroughly  investigated in \cite{Mandelbaum2005}.
We measure the stacked WL excess surface mass density profile, $\Delta\Sigma(r)$ (\autoref{eq:DSigma}), in 16 logarithmically-spaced radial bins in the range 0.3--50~\mpch. We present the resulting profiles  in \autoref{fig:Sigma} for both the CC/NCC (left, circles and triangles, respectively) and the strong CC/NCC (right) samples. The covariance is derived using the jackknife method, dividing the sample into 83 equal area bins. The figure demonstrates that the two subsamples have very similar mass profiles.

We fit for both the one- and two-halo terms simultaneously, setting flat priors on the mass, \m200, concentration, $c_{\rm 200m}$, and halo bias $b$. We set restrictive gaussian priors on the miscentering parameters, around the nominal values presented in \cite{Simet2017}. We do this since the profiles are not  well constrained at the center, $r<300$~kpc, and miscentering is highly degenerate with the concentration parameter. For the same reason, we do not fit for a central stellar mass contribution from the BCG.
The fitted masses are listed for each cluster sample in \autoref{tab:samples}.
We find overall consistent masses for the CC/NCC samples, $M_{\rm CC} = 1.99^{+0.21}_{-0.19}\mhunit$, and  $M_{\rm NCC} = 2.10^{+0.18}_{-0.16}\mhunit$. 
The strong CC/NCC also show similarly consistent masses,  $M_{\rm stCC} = 1.99^{+0.45}_{-0.40}\mhunit$,  and  $M_{\rm stNCC} = 2.11^{+0.23}_{-0.21}\mhunit$.
For both cases, mean masses are consistent within 1$\sigma$.

As presented in \autoref{sec:WLmethod}, we can calculate the expected mass-only bias (\autoref{eq:bmlens}). Using the  \cite{Tinker2010} halo bias model provided in the python {\sc colossus} toolkit \citep{Diemer2018}, we translate each lensing mass in the MCMC chains to bias. We then derive the relative bias by dividing each of NCC bias values with each of the CC  bias values. 
The expected bias ratio of CC and NCC clusters is $\bmlens = 1.05^{+0.07}_{-0.06}$. For the strong CC and strong NCC samples, the expected relative bias is $\bmlens = 1.05^{+0.13}_{-0.11}$. Therefore, based on the masses, the LSS around NCC is not expected to be significantly  more clustered  than around CC clusters
 ($\lesssim5\%$).
 These results are summarized in the first row of \autoref{tab:results}.
 
In comparison, the linear biases directly estimated from the two-halo fit show a ratio that is statistically consistent with that   expected from the lensing masses, though the central value is about $1\sigma$ below. In the case of CC/NCC,  we measure $b_{\rm CC} = 2.87\pm{0.39}$, and  $b_{\rm NCC} = 2.47^{+0.42}_{-0.43}$. 
As defined in \autoref{eq:blens}, this translates to a relative bias of $\blens = 0.86\pm{0.20}$. For the strong CC/NCC samples, we find $b_{\rm stCC} = 2.75\pm{0.78}$, and  $b_{\rm stNCC} = 2.10^{+0.49}_{-0.47}$, i.e., a relative bias of $\blens  = 0.89^{+0.67}_{-0.54}$.  The two-halo bias results are summarized in the second row of \autoref{tab:results}. We will  quantify   the significance and interpretation of these different results in \autoref{subsec:assbias}, together with the two-point clustering results presented next.

\begin{deluxetable}{@{\extracolsep{4pt}}lcccc}
\tablecaption{Results of the bias analysis from different methodologies\label{tab:results} }
\tablewidth{0pt}
\tablehead{
\colhead{} &  
\multicolumn{2}{c}{CC/NCC} & \multicolumn{2}{c}{Strong CC/NCC}\\
\cline{2-3} \cline{4-5} 
\colhead{Method}& \colhead{\b} & \colhead{$f$} & \colhead{\b} & \colhead{$f$} 
} 
\startdata
Lensing\\
1-halo (expected) & $1.05_{-0.06}^{+0.07}$ & $1$ &          $1.05_{-0.11}^{+0.13}$ & $1$  \\
2-halo & $0.86_{-0.20}^{+0.20}$ & $0.83_{-0.20}^{+0.20}$ &          $0.89_{-0.54}^{+0.67}$ & $0.86_{-0.52}^{+0.66}$  \\
\hline
Clustering\\
\ccf & $1.01_{-0.03}^{+0.03}$ & $0.97_{-0.06}^{+0.07}$ &          $0.98_{-0.06}^{+0.06}$ & $0.94_{-0.11}^{+0.13}$  \\
\ccf, projected & $0.97_{-0.06}^{+0.06}$ & $0.93_{-0.08}^{+0.08}$ &          $0.82_{-0.10}^{+0.10}$ & $0.79_{-0.12}^{+0.13}$  \\
\acf & $1.02_{-0.06}^{+0.06}$ & $0.98_{-0.08}^{+0.08}$ &          $0.97_{-0.22}^{+0.22}$ & $0.93_{-0.23}^{+0.25}$  \\
\enddata
\tablecomments{
$\b = b_\mathrm{NCC}/ b_\mathrm{CC}$ measures the level of relative bias between the NCC and CC clusters.
$f=\b/\bmlens$ measures the level of assembly bias. 
Median and 68\% confidence bounds are quoted.
}
\end{deluxetable}

\subsection{Clustering}
\label{subsec:bias}

\begin{figure*}
	\includegraphics[width=\textwidth,clip]{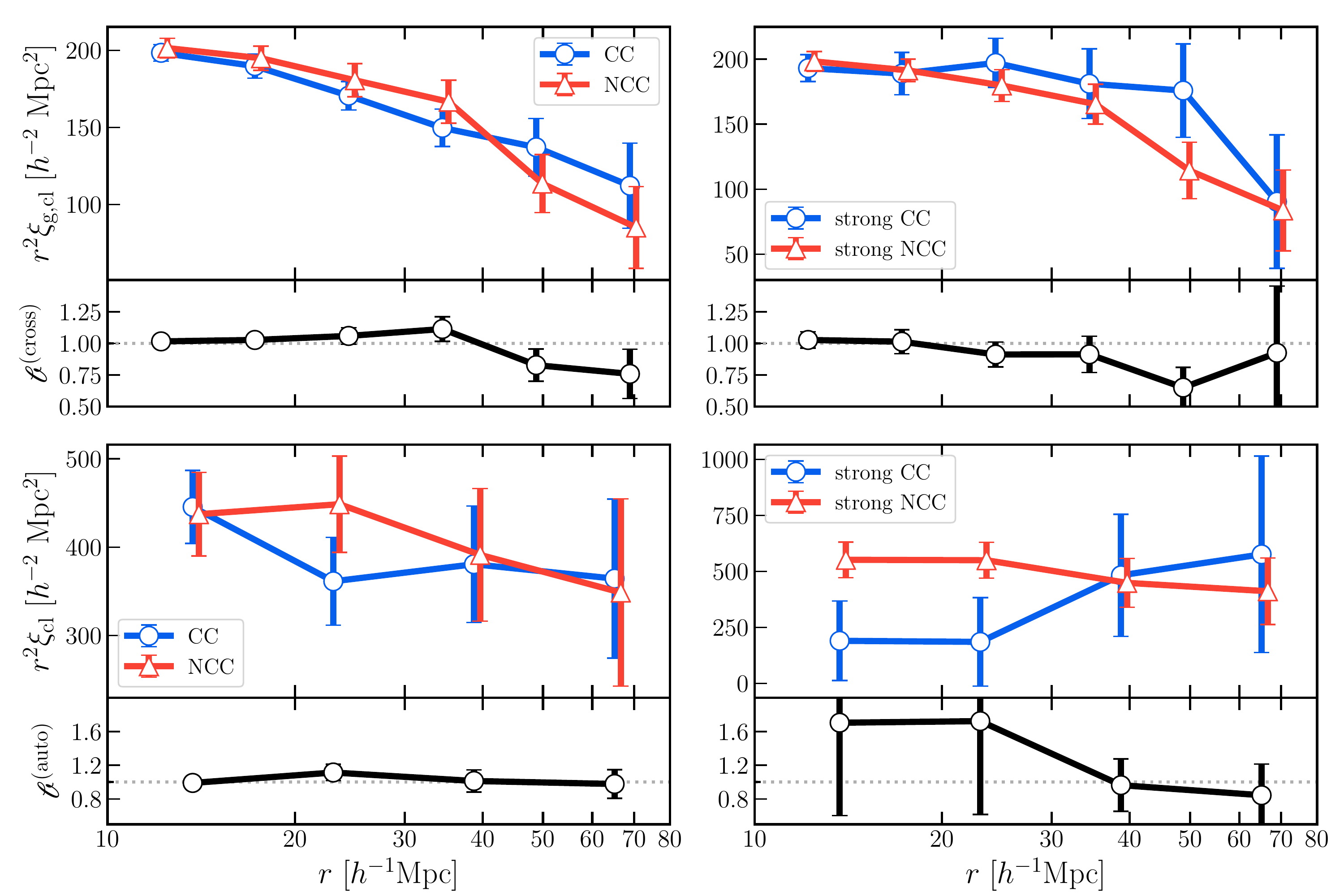}
	\caption{Each upper panel in the plots shows the real-space two-point correlation functions for the CC (blue) and and NCC (red), while the bottom panels show the relative bias (black) derived from the ratio of the CC/NCC correlation functions above it, as a function of comoving separation, $r$. Left plots show the results for the median-divided CC/NCC samples, whereas right plots show the strong CC/NCC samples  (see \autoref{tab:samples} for definitions). 
Top and bottom plots use different clustering methodologies.
Top: using the CCFs between  galaxies and  clusters, \ccf(r). 
Bottom: using the ACFs of the cluster samples, \acf(r).
Overall, the results are consistent between the different methodologies. The CCF methodology, however, yields tighter constraints than the ACF due to the larger galaxy sample employed.
}	
	\label{fig:xcorr}
\end{figure*}

\begin{figure*}[bt]
	\includegraphics[width=\textwidth]{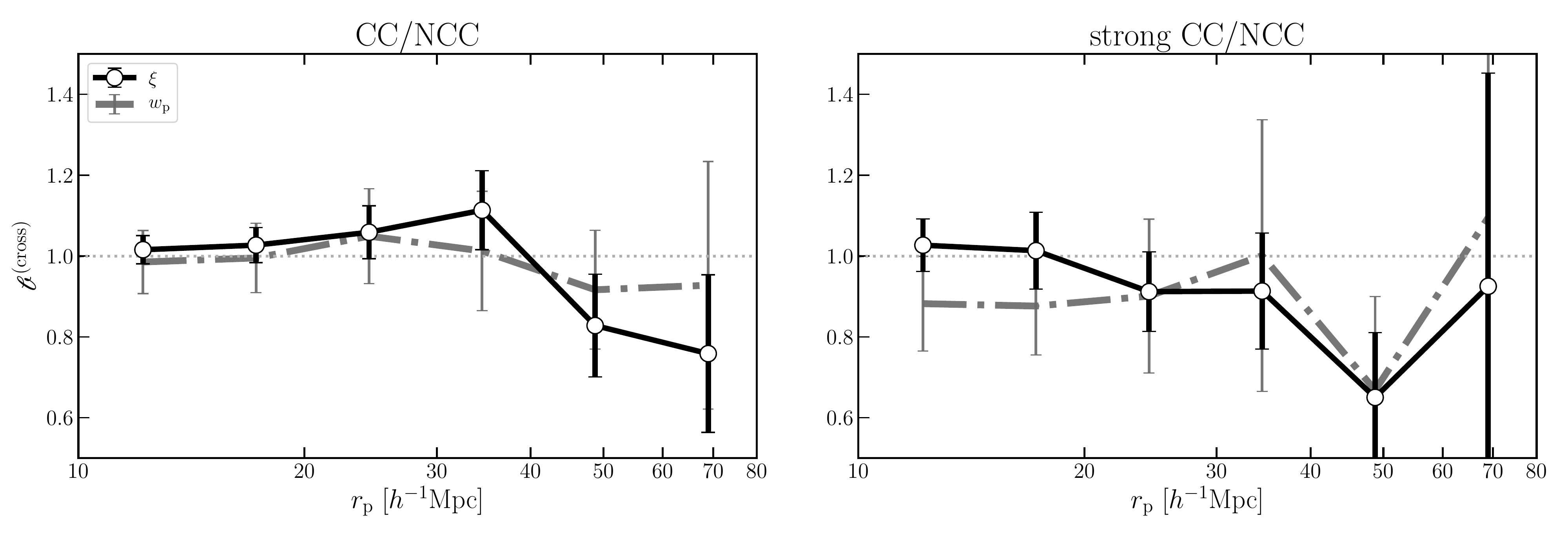}
	\caption{Relative bias  from cross-correlation as a function of separation, but calculated using the {\it projected} CCF, $w_{\rm p}(r_{\rm p})$ (gray dash-dotted line).  The signal is consistent with that derived from the real-space CCF (black, same as the black solid lines in the top plots of  \autoref{fig:xcorr}).}

	\label{fig:wp}
\end{figure*}



Here we measure the relative bias from the clustering profiles on large-scales using the two-point correlation functions as 
defined in \autoref{sec:method}.
We make use of the public code {\sc corrfunc} \citep{Sinha2017}, which relies on the \cite{LS93} estimator, to calculate all the two-point correlation functions. We first compute the CCF of each cluster subsample (CC, NCC, strong CC and strong NCC) with the LOWZ galaxies. In order to avoid redshift-space distortions that affect galaxies infalling into halos, we need to compute the CCF outside the cluster one-halo regime. We evaluate this scale by computing the velocity dispersion of each cluster subsample from the peculiar velocities of LOWZ galaxies inside 1.5~\mpch{} of a nearby cluster halo. The CC and NCC samples show $\Delta V \simeq 700$~km/s, corresponding roughly to a virial radius of 1~\mpch. We furthermore compute the  two-dimensional CCF as a function of projected ($r_p$) and line-of sight ($\pi$) distance, $\xi(r_p,\pi$). 
Finger-of-god effects are evident up to scales $\lesssim10$~\mpch. We therefore choose to compute the real space CCF (\autoref{eq:ccf}) in  six logarithmic bins in the range 10--80~\mpch. Throughout, the full covariance matrix is derived using the jackknife method for each of the correlation functions and the bias profiles,  by diving the SDSS/\RM{} area into 192 equal area regions. The results are presented in the top panels of \autoref{fig:xcorr}. The upper panel of the left plot shows the CCFs, $\ccf(r)$, of the CC (blue) and NCC (red) clusters, and the bottom panel shows the bias, $\bcross(r)$ (black), derived from the ratio of the two CCFs  (\autoref{eq:bias_cross}). The right panel shows the same for the strong CC and strong NCC cluster subsamples. To estimate the mean bias level we fit $\b(r)$ with a constant  model taking into account the  covariance between  scales. The median relative bias, $\bcross$, and its uncertainty, are summarized in \autoref{tab:results}. Apparently, there is no significant difference between the clustering around CC versus NCC,  
$\bcross = 1.01\pm0.03$. 
The bias for strong CC/NCC is also non-significant,  
$\bcross= 0.98\pm0.06$. 

Another common methodology is to calculate the ACF of the clusters themselves,  $\xi_{\rm cl}(r)$, given by \autoref{eq:acf}, where the relative bias equals the square-root of the ratio of the ACFs (\autoref{eq:bias_auto}). Naturally, this methodology  yields larger errors, but may serve as a semi-independent check of the bias, since the LOWZ galaxy sample is not included here. We present the results for the CC/NCC and strong CC/NCC samples in the bottom plots of \autoref{fig:xcorr}.
The ACF  also show a bias fully consistent with the expected level,  $\bauto = 1.02\pm0.06$ for the CC/NCC, and   $\bauto =0.97\pm0.22$  for the strong CC/NCC. The results are summarized in the last row of \autoref{tab:results}.
We will  compare the clustering results with those derived from lensing and interpret them in the context of assembly bias in \autoref{subsec:assbias}.

Lastly, to validate that the  real-space CCF are not affected by residual redshift-space distortions (RSD) effects, we also compute the projected CCF, $\wp(r_{\rm p})$, integrated along the line of sight to $\pi=100$~\mpch. Such integration removes any issues due to RSD. The results, shown as dashed lines in \autoref{fig:wp}, are  consistent with those derived from the real-space CCF, $\ccf(r)$.

\section{Discussion}
\label{sec:discussion}

\subsection{Significance of Assembly Bias}
\label{subsec:assbias}

\begin{figure*}[bt]
\includegraphics[width=\textwidth,clip]{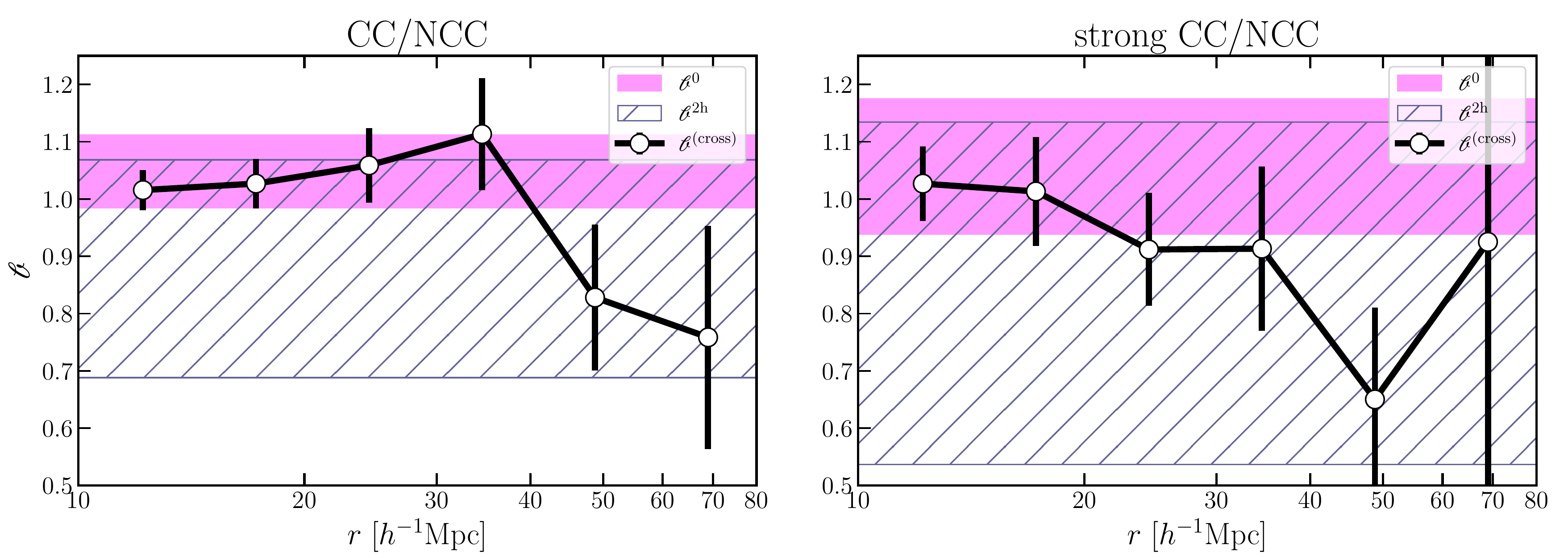}

	\caption{Comparison of the  relative bias measured with cross-correlation, \bcross$(r)$ (black points, same as in top panels of \autoref{fig:xcorr}) and with lensing, \blens (68\% confidence region; hatched). These two are compared with the level of expected mass-only relative bias \bmlens (68\% confidence region; pink), calculated using  the \cite{Tinker2010} halo bias model with the WL-derived masses as input. The two measurements are consistent within $\lesssim1\sigma$ with the expected mass-only bias level, for both the CC/NCC and strong CC/NCC samples, i.e., showing no evidence for assembly bias between CC and NCC.}	
	\label{fig:b_lens}
\end{figure*}

We now explore the level of assembly bias derived from each methodology by comparing it with the expected mass-only bias level, \bmlens, derived from the fitted lensing masses.
To visualize these, in \autoref{fig:b_lens} we plot the fiducial bias profile measured from the ratio of CCFs, \bcross{} (black; same as solid black line in \autoref{fig:xcorr}). We overlay the lensing-measured relative bias using the fitted bias parameters, \blens{} (hatched, 68\% confidence region), and finally the expected mass-only bias derived by assuming a \cite{Tinker2010} halo bias model with the WL mean masses as input, \bmlens{}  (pink, 68\% confidence region). It appears that \bcross agrees well with the level derived from the lensing two-halo term, \blens, though the constraints on the latter are   poor ($\sim20-60\%$ confidence level). 
On the other hand, both measured biases, \bcross{} and  \blens{},  are about 1$\sigma$ below the expected level, \bmlens{}.

To better quantify if the measured bias significantly differs from the expected mass-only bias \bmlens, we fit each measured dataset (\blens, \bcross(r), projected $\bcross(r_{\rm p})$ and $\bauto(r)$) with the model described in \autoref{eq:f}. For all the clustering results, which were determined as a function of scale, we take into account the full covariance matrix in our likelihood function. Since our parameter of interest is $f$,  we set a flat (uninformative) prior on $f$. We set a tight lognormal prior on \bmlens{} dictated by the median and standard deviation of $\log(\bmlens)$ and then marginalize over it. The posterior values for $f$ are given in \autoref{tab:results} next to each measured \b.

As can be seen, all values of $f$ are consistent with the null hypothesis, $f=1$, up to 1--2$\sigma$. The projected \bcross{} measured for the strong CC/NCC samples shows the highest offset, $f=0.79\pm0.12$ but is still below the $2\sigma$ level. 
We therefore conclude, based on the \RM{} cluster sample, that we find no indication of different assembly histories for CC and NCC clusters to within 6\% uncertainty.

\subsection{Comparison with Simulations}
\label{subsec:sims}

In  simulations, 
assembly bias on cluster scales
has been shown to exist in practically all definitions of halo age
 \citep{Chue2018}, although some proxies appear to be noisier than others \citep{Mao2018}. 
Recently, combining several N-body simulations covering a wider mass range up to $10^{14.5} h^{-1}\,M_\odot$, \cite{Sato-Polito2018} find different secondary indicators of assembly histories yield somewhat different results. For example, if separating halos by their age, no assembly bias is detected above \mhunit. On the other hand, if separating halos by concentration or spin, at the high-mass end, the difference between the upper and lower quartiles can reach up to a factor of 2.

There is no equivalent in cosmological simulations for the physical distinction we are studying, i.e., the cooling phase of the cluster core gas. This is since baryonic effects, especially considering the large range in scales involved, are currently hard to include \citep[although see][]{Rasia2015,Barnes2018}. Therefore, any statements about the levels found in simulations are not necessary applicable to our study. Nonetheless, we may consider the above range of bias values found in simulations as a guideline to the  degree of assembly bias expected in general. With that in mind, it is safe to conclude we find no significant level of assembly bias in our samples.

Given this measurement alone, we may not identify the specific formation model. Whether  cores are created early on and left undisturbed, or whether later periodic CC formation and distruption and in play,  we find those to be decoupled from the external environment on large scales. One may argue this is not surprising since the cores are embedded in self-similar envelopes \citep{McDonald2017}. 
It may, in turn, be that even though their large-scale modes appear the same, small-scale fluctuations result in different initial conditions at the halo regime, e.g., NCC may be the result of a lower amplitude fluctuation with higher substructure. It would thus be  interesting to compare the small-scale clustering of CC and NCC clusters at low and high redshift.

\section{Summary \& Conclusions}
\label{sec:summary}
In this paper we tested the level of assembly bias between different cluster samples, separated by a physically-motivated criteria: the level of cooling in their cores. We differentiate between CC and NCC clusters solely based on the BCG \ha{} luminosity as a cooling indicator. We draw samples of thousands of CC and NCC clusters from the SDSS/DR8 \RM{} catalog to achieve a statistically significant measurement of the weak assembly bias effect. 
We performed a weak lensing analysis measuring the cluster density profiles to investigate their mean mass and two-halo linear bias properties.
Furthermore, we applied a complementary and novel methodology, cross-correlating  the cluster samples of  with an even larger sample of hundreds of thousands of galaxies from the LOWZ sample, to gain better statistical precision on the bias. This method provides information on the large-scale environments of clusters who have apparently different characteristics, which in turn  provide insight into their formation history. 

From WL we found  the CC and NCC samples  to  have comparable mean masses,    $\m200 \approx2\times\mhunit$. From the mean WL masses, we quantified the expected level of mass-only bias to be  $\bmlens = 1.05^{+0.07}_{-0.06}$ for the CC/NCC samples, and  $\bmlens = 1.05^{+0.13}_{-0.11}$ for the strong CC/NCC samples. We then quantified the departure from the null hypothesis with the assembly bias fraction, $f$, and fitted the measured bias to constrain this parameter.
From the lensing two-halo measured bias, \blens, we found no indication of assembly bias, with $f=0.83\pm0.20$ for CC/NCC and $f=0.86^{+0.66}_{-0.52}$ for the strong CC/NCC. Since the size of the error is approximately at or greater than the expected level of the effect, $\sim20-70\%$, for the current sample size lensing alone lacks statistical constraining power to probe assembly bias.

From the more sensitive CCF methodology, we fit \bcross and found $f=0.97^{+0.07}_{-0.06}$ for the CC/NCC samples and $f=0.94^{+0.13}_{-0.11}$ for the strong CC/NCC samples. We also applied the more traditional ACF methodology using the clusters alone, 
and found $f=0.98\pm0.08$ for CC/NCC and $f=0.93^{+0.25}_{-0.23}$ for the strong CC/NCC, though this methodology is statistically weaker, with  similar constraining power as of the lensing methodology.
Therefore, within $\sim1\sigma$, both subsets using both methodologies agree with no assembly bias.

It is important to note that although we have  expanded our  analysis to make use of thousands of clusters compared to dozens in \citetalias{Medezinski2017}, the \RM{} sample is inherently biased against heavily star-forming cluster cores \citep{Rykoff2016}. Indeed, our strong CC sample is limited to $\sim$500 clusters only. A future study would benefit from a less biased sample, e.g., not relying on a red-sequence finder \citep[e.g.,][]{Soares-Santos2011,Wen2012}, or one that makes use of future multi-narrowband surveys (e.g.,  SPHEREx; \citealt{Dore2016}).

We conclude, based on the results presented in this analysis, that any observed differences between CC and NCC clusters are not inherited from different large-scale environments. In turn, the assignment of the cluster cooling phase may result from different small-scale clustering. Combined with the reported mild evolution in CC properties, one possible  solution is that the local gas properties in the protocluster core  determine the subsequent creation of an AGN feedback loop. 
Alternatively,  local merger activity could create and destroy CC periodically  but in a way that mimics constant mean population properties over time.

While our study cannot confirm either proposed picture, studying clusters at the onset of formation, which appears to be as high as $z\sim2$, can shed direct light on the process of core formation.
We are only now at a stage where the evolution  of clusters can be studied statistically with large ensembles, both in observations \citep{McDonald2017,McDonald2018} and in cosmological simulations \citep{Rasia2015,Barnes2018}. 
Such large-scale simulations are  at the infancy of incorporating complex AGN feedback and feeding physics via subgrid models, and require better physically-motivated (rather than fine-tuned) schemes in order to achieve robustly predictive results for the CC-NCC formation \citep[e.g.,][]{Gaspari2017}.
LSST and WFIRST will revolutionize our understanding of  cluster formation and evolution, as they will enable us to study the galaxy populations in $z\lesssim2$ clusters in the thousands instead of the current dozens. CMB-S4 will both detect  clusters up to $z\sim2$ through the redshift-independent SZ effect and inform on the gas properties at the epoch of clusters formation.
Our most pressing challenge is matching those promising optical/IR surveys with the complementary high-resolution spectral and X-ray surveys of cluster cores (e.g.,   {\it Athena}; \citealt{Ettori2013}) that will facilitate better understanding of the initial buildup and the precise impact of baryonic physics on cluster formation.


\acknowledgments
We acknowledge insightful discussions with Michael Strauss, Crist\'obal Sif\'on, Neta Bahcall, Andrew Hearin, Andy Goulding, Mathew Madhavacheril, Andrina Nicola, Jim Gunn and Jenny Greene.
M.G. is supported by the Lyman Spitzer Jr. Fellowship (Princeton University) and by NASA Chandra GO7-18121X.
Funding for SDSS-III has been provided by the Alfred P. Sloan Foundation, the Participating Institutions, the National Science Foundation, and the U.S. Department of Energy Office of Science. The SDSS-III web site is http://www.sdss3.org/.
SDSS-III is managed by the Astrophysical Research Consortium for the Participating Institutions of the SDSS-III Collaboration including the University of Arizona, the Brazilian Participation Group, Brookhaven National Laboratory, Carnegie Mellon University, University of Florida, the French Participation Group, the German Participation Group, Harvard University, the Instituto de Astrofisica de Canarias, the Michigan State/Notre Dame/JINA Participation Group, Johns Hopkins University, Lawrence Berkeley National Laboratory, Max Planck Institute for Astrophysics, Max Planck Institute for Extraterrestrial Physics, New Mexico State University, New York University, Ohio State University, Pennsylvania State University, University of Portsmouth, Princeton University, the Spanish Participation Group, University of Tokyo, University of Utah, Vanderbilt University, University of Virginia, University of Washington, and Yale University.
The work reported on in this paper was substantially performed at the TIGRESS high performance computer center at Princeton University which is jointly supported by the Princeton Institute for Computational Science and Engineering and the Princeton University Office of Information Technology's Research Computing department.


\appendix


\begin{figure*}[tb]
\includegraphics[width=0.5\textwidth,clip]{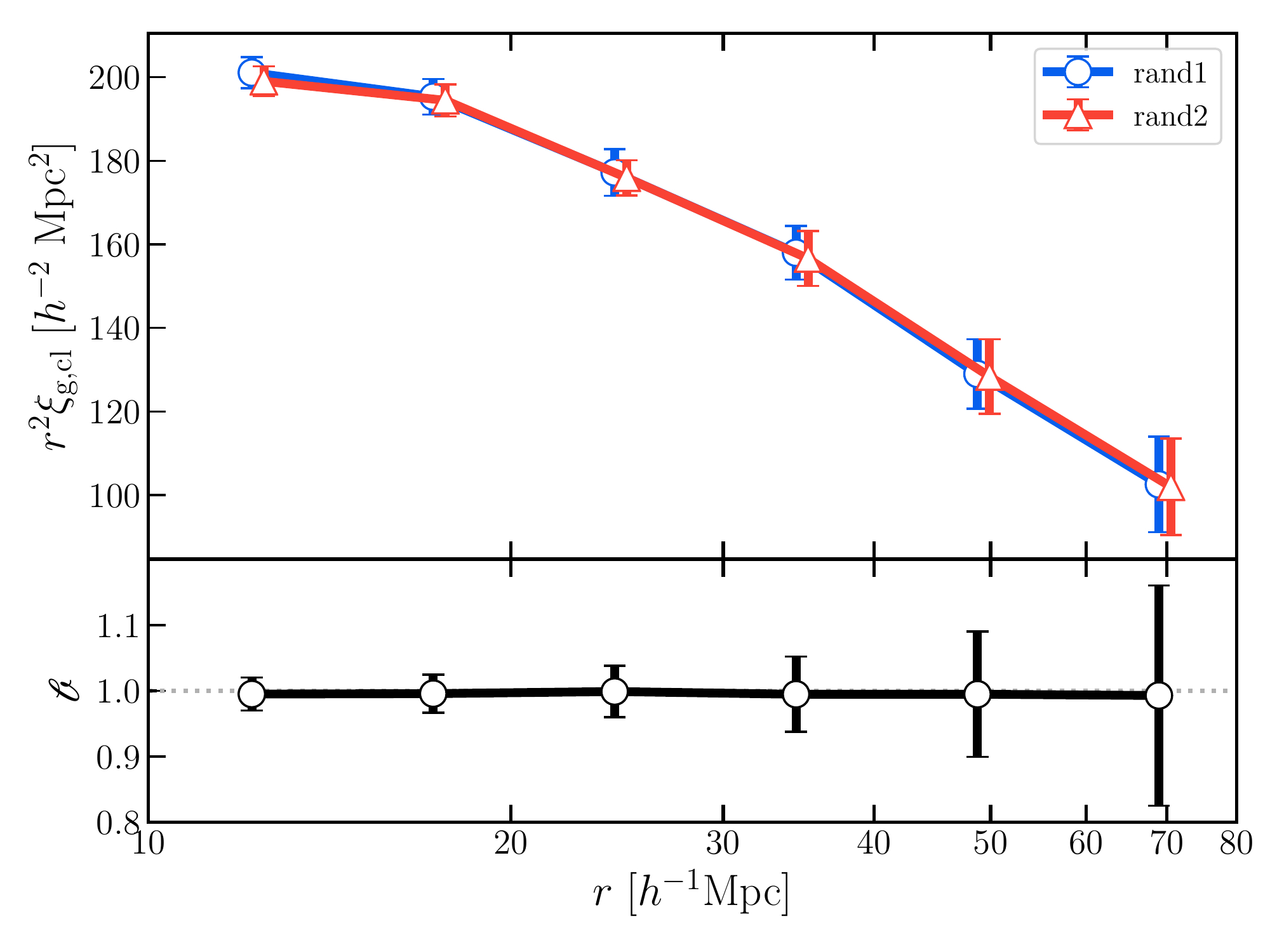}
	\includegraphics[width=0.5\textwidth,clip]{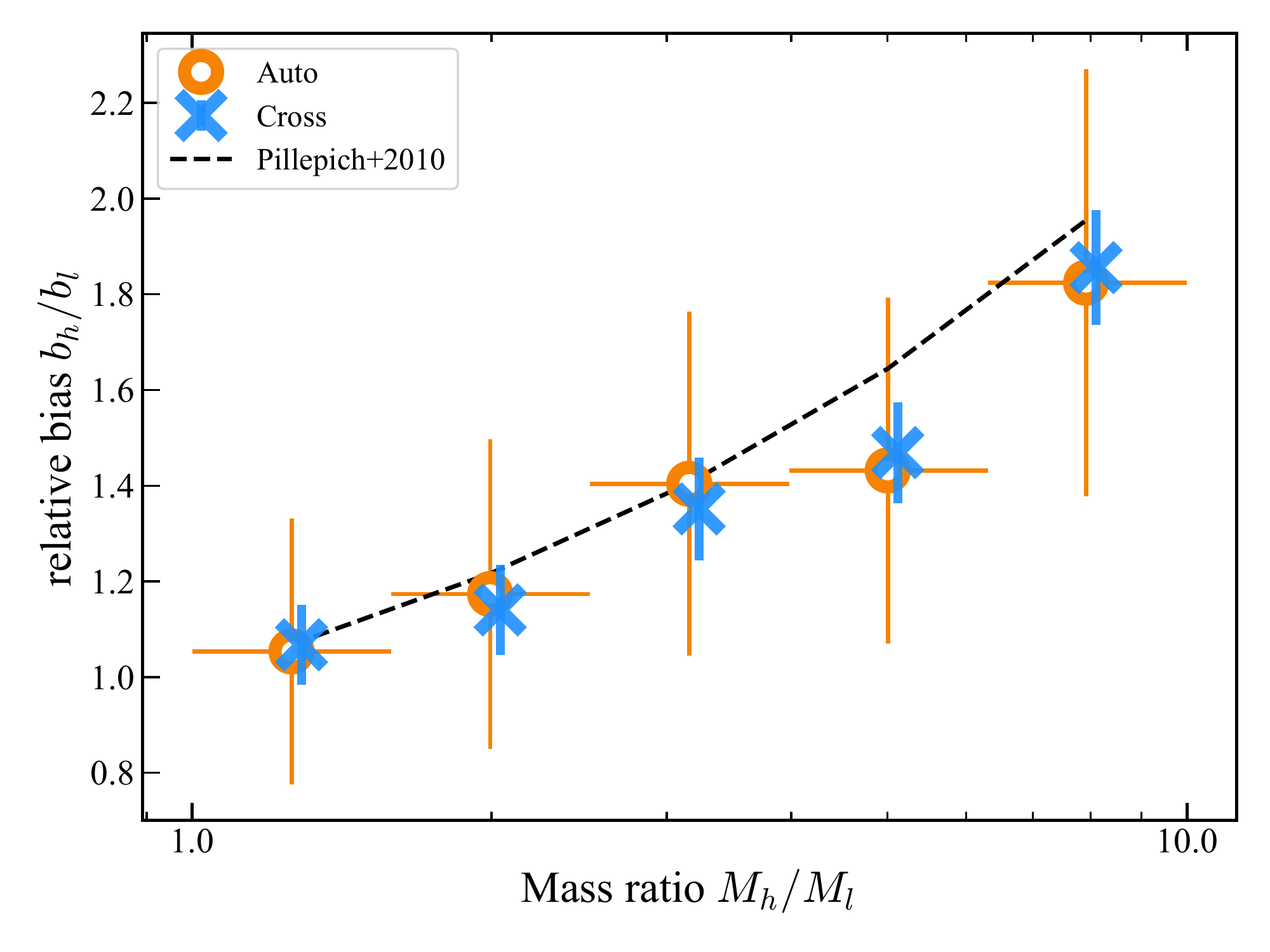}

	\caption{Left: Results from the null test. We randomly select subsamples from the parent \RM{} catalog and calculate the CCFs and relative bias, clearly validating the null hypothesis: $b=1$  between two random subsamples.
	Right: Results from the mass bias test. The theoretical relation between relative halo mass bias and the mass ratio is given by the black dashed line, based on \cite{Pillepich2010} simulations. Orange circles and blue crosses show the  relative bias in each  mass ratio bin measured from simulated MICE clusters and galaxies using the ACF and CCF, respectively. The ACF and CCF methodologies give consistent results, and both are consistent with the expected level.}
	\label{fig:test}
\end{figure*}

\section{Null test}
\label{app:null}
We  perform a null test to validate that our CCF methodology yields no bias  ($\b=1$) when selecting cluster subsamples randomly.
From the parent \RM{} sample, we  select 100 random subsamples each of the same size as our CC/NCC samples ($N=3000$). We preform the same CCF analysis as described in \autoref{subsec:bias}, cross-correlating each random subsample with the LOWZ catalog. We divide each of the CCFs of the first 50 subsamples (`rand1') with each of the last 50 subsamples' (`rand2') CCFs to get $50^2$ relaltive bias realizations.  The results are presented in the \autoref{fig:test} (left). Errors represent the scatter from the random sampling. These errors also take into account both statistical noise and systematics due to the assumption that equal richness distributions represent equal masses. These errors are comparable to the jackknife errors which we use as our fiducial errors.


\section{Mass bias test}
\label{app:massbias}
In this section we test that our CCF methodology retrieves the expected mass bias.
For this test, we use the Marenostrum Institut de Cie\'ncies de l'Espai (MICE)  Grand Challenge (MICE-GC) simulations \citep{Fosalba2015,Fosalba2015a,Crocce2015,Carretero2015,Hoffmann2015}, for which the halo mass is known. We define clusters as having $M_{\rm h}>10^{13}M_\sun$.
We randomly select 20 bootstrap realizations of cluster samples each of size $N=3000$, in each of the six mass bins. We cross-correlate each cluster sample with a mock galaxy sample (selected to mimic the LOWZ sample in terms of color and redshift distribution) over the entire radial range (10 to 80~\mpch, as for our data), and divide each of the CCFs of the five highest-mass bins  with the CCF of the lowest mass bin. This gives us a test of our CCF methodology: the ratio should give the relative mass bias,  \bcross. Since we have 20 mock cluster catalogs in each mass bin, we have overall 400 realizations of the relative bias  in each mass ratio bin. We take the scatter  as the error on the relative bias.
We follow the same exercise for the ACF methodology, measuring \bauto.
We plot the expected (black; using the \cite{Pillepich2010} model) and the derived   \bcross{} (blue crosses) and \bauto{} (orange circles) averaged over all scales as a function of  mass ratio in  \autoref{fig:test} (right). As can be seen, the two methodologies are completely consistent with each other. As expected, the CCF methodology yields much smaller uncertainties than ACF. The CCF-derived \bcross{} is slightly below the expected values found by  \cite{Pillepich2010}, but only at high mass ratios (our study is at a mass ratio of $\lesssim1.2$, as determined in \autoref{subsec:WL}).  We also note that different approaches adopted by different simulation sets may cause this small effect.


\bibliography{/Users/elinor/Dropbox/Documents/Elinor.bib}

\end{document}